\documentclass{ws-procs9x6-cpt16}
\begin{document}

\newcommand{\refeq}[1]{(\ref{#1})}
\def\etal {{\it et al.}}
%any other macros go here 

\title{Musings on Lorentz Violation Given the \\
Recent Gravitational-Wave Observations of \\
Coalescing Binary Black Holes}

\author{Nicol\'as Yunes}

\address{eXtreme Gravity Institute, Department of Physics, Montana State University\\
Bozeman, MT 59717, USA}

\begin{abstract}
The recent observation of gravitational waves by the LIGO/Virgo collaboration provides a unique opportunity to probe the \emph{extreme gravity} of coalescing binary black holes. In this regime, the gravitational interaction is not only strong, but the spacetime curvature is large, characteristic velocities are a non-negligible fraction of the speed of light, and the time scale on which the curvature and gravity change is small. This contribution discusses some consequences of these observations on modifications to General Relativity, with a special emphasis on Lorentz-violating theories. 
\end{abstract}

\bodymatter

\section{The first probe of extreme gravity}
In the Fall of 2016, the LIGO/Virgo Collaboration (LVC) detected the gravitational waves produced in the late inspiral and merger of two binary black-hole systems.\cite{Abbott:2016blz} These signals were found to be consistent with that predicted by General Relativity (GR) for a compact system with masses of $(36,29) M_{\odot}$ and $(14,8) M_{\odot}$. The signals were so ``loud'' that not only was the associated statistical $\sigma$ in excess of five, but the first event's wave oscillations could even be seen by eye in the data stream.  

The black-hole binary systems that produced these gravitational waves are unique because they sample a regime that had evaded probing thus far: \emph{the extreme gravity regime}.\cite{Yunes:2013dva} Extreme gravity refers to regions of space and time in which gravity is strong (relative to the solar system), characteristic velocities are a large fraction of the speed of light, the curvature of spacetime is large, and the time scale on which gravity changes is very short. The binary systems the LVC observed consisted of black holes, which are intrinsically strongly gravitating objects that curve spacetime dramatically, accelerating from speeds of $10\%$ to $50\%$ the speed of light in less than one second (see, e.g., Figs.\ 1-3 in Ref.\ \refcite{Yunes:2016jcc}).

\section{Raising the bar that modified gravity must pass}
Until the LVC observations, modified gravity\cite{Berti:2015itd} could get away with claiming viability by satisfying solar system, binary pulsar and cosmological constraints. This is, by no means, a small feat. Over the past century, a plethora of tests have been developed, which could place an array of constraints on such modifications.\cite{Will:2005va}  For example, Lorentz violation typically modifies the Einstein equivalence principle of GR, which has been tested to incredible precision with Michelson-Morley type experiments, atom-clock anisotropy experiments, observations of gamma-ray bursts from blazars, reaction rates at particle accelerators, birefringence effects in the propagation of photons, and neutrino oscillations.\cite{Will:2014bqa}

The direct observation of gravitational waves from the coalescence of binary black holes, however, raises the bar that modified theories must pass to incredibly high levels.\cite{TheLIGOScientific:2016src,Yunes:2016jcc} Let us dissect a minimal set of requirements that, in my opinion, modified theories must now satisfy. First, very compact, highly spinning and very massive objects must exist in the theory \emph{and} they must be stable. The compact objects that the LVC observed did not magically appear just during the LVC observations; rather, each object probably formed many millions of years before the merger and, during this time, it remained stable to perturbations. Second, the theory must predict the existence of some energy sink that forces a binary system to decay, to \emph{inspiral}, at the observed rate (predominantly quadrupolarly). Third, the theory must allow for such compact objects to collide and form a larger compact object that settles down to a stationary configuration in roughly one light-crossing time. This implies the compact objects must have an immense effective viscosity to dissipate any deformations efficiently. Fourth, the remnant compact object must possess some type of light ring (or photon sphere) that vibrates appropriately to produce the dominant ringdown signal observed by the LVC. Fifth, during such a coalescence, the theory must predict the existence of gravitational waves (oscillations in the metric far away from the source) that propagate at some speed and weakly interact with any intervening matter. Such waves must induce a predominantly quadrupolar response on detectors (on the travel time of photons). Sixth, the dispersion relation of gravitational waves must be consistent with the standard relation for a massless gauge boson.

\section{Looking into the future}

Do Lorentz-violating modifications satisfy the requirements presented above?  This is not clear, but not because we know they do not. Rather, the theory of Lorentz-violating gravity has not advanced sufficiently to answer this question. Let us consider Einstein-Aether theory as a concrete example. This theory modifies GR through the inclusion of dynamical vector fields that select preferred directions in spacetime, and thus, violate Lorentz invariance.\cite{Jacobson:2008aj} Although recently black-hole solutions have been found numerically in this theory,\cite{Barausse:2015frm} a stability analysis has not yet been carried out. Similarly, although a leading-order post-Newtonian analysis has been carried out for neutron-star binaries,\cite{Hansen:2014ewa} the corresponding analysis for black-hole binaries is currently missing. Without this information, it is simply impossible to compare quantitatively the predictions of Einstein-Aether theory to the LVC observations (only qualitative comparisons are possible\cite{Yunes:2016jcc}). This, however, is not due to laziness by the theory community, but rather due to the difficulty inherent in the modified field equations that describe compact binaries.

This situation is even worse in generic Lorentz-violating models, such as the Standard-Model Extension (SME). The latter is an \emph{effective theory} that modifies the Einstein-Hilbert action by adding all possible terms, produced by contractions of scalar, vector and tensor fields and their covariant derivatives, multiplied by certain Lorentz-violating fields.\cite{Kostelecky:2003fs} To address the requirements described earlier, the SME needs to be \emph{nonlinearly} studied in the gravity sector to determine, for example, whether black holes exist in the SME and what their properties are. One could attempt to construct the gravitational field outside a mass distribution as a perturbative series in the ratio of the mass to its radius (a weak-field expansion). But to obtain a black-hole solution, one would then have to take the limit as the radius tends to the mass, forcing the series to typically diverge. Without an appropriate resummation technique, it is not clear that such a perturbative approach will lead to the correct black-hole solutions of the theory. 

Another important issue is the modeling of the merger in such effective gravity theories. Typically, given a particular theory, one can carry out a 3+1 decomposition of the field equations and numerically solve them to obtain the merger.\cite{Lehner:2014asa} Effective gravity theories, however, typically include higher derivatives in the field equations, which change the characteristic structure of the system and are likely to be unstable. Of course, the problem here is that an effective theory is not an exact theory, and in particular, effective theories contain high-order remainders that must be removed (e.g., via order reduction in perturbative solutions or via numerical filtering). Currently, there are no implementations of such order-reduction ideas in numerical simulations of compact-object coalescences.  

So should we just give up? Of course not. The LVC has discovered the tip of the iceberg, with many more gravitational-wave observations to come. The bar I described earlier will continue to be raised, which means that we should be able to constrain stringently more and more interactions in extreme gravity (including Lorentz-violating effects). Given the status of the field, we are today limited to qualitative tests that, e.g., only include modifications in the propagation of gravitational waves;\cite{Kostelecky:2016kfm,Yunes:2016jcc} such constraints are likely to be overly conservative, i.e., knowledge of Lorentz-violating black holes and the dynamics of their merger are likely to lead to interesting constraints. This will require a consorted effort of gravitational-wave theorists, particle theorists, and numerical and analytical relativists. 

\section*{Acknowledgments}
These musings are based on Ref.\ \refcite{Yunes:2016jcc}, which will soon be published in Physical Review D and are a summary of a talk given at the CPT'16 conference at Indiana University. I would like to thank Indiana University and the organizers of this meeting for their hospitality duration the duration of the meeting. I would also like to thank Kent Yagi and Frans Pretorius for a careful reading of this manuscript and for many discussions. This work was supported by the NSF CAREER Grant PHY-1250636.

\end{document}